\documentclass[pre,twocolumn,showpacs,preprintnumbers,reprint,amsmath,
               amssymb,superscriptaddress]{revtex4-1}
                
\usepackage{graphicx} 
\usepackage{dcolumn}  
\usepackage{bm}       
\usepackage{pstricks}

\begin{document}

\newcommand{\bea}{\begin{eqnarray}}
\newcommand{\eea}{  \end{eqnarray}}
\newcommand{\bit}{\begin{itemize}}
\newcommand{\eit}{  \end{itemize}}

\newcommand{\be}{\begin{equation}}
\newcommand{\ee}{\end{equation}}
\newcommand{\ra}{\rangle}
\newcommand{\la}{\langle}
\newcommand{\U}{\widetilde{U}}

\newrgbcolor{fbkblue}{0.2304 0.3476  0.5937}
\newrgbcolor{pksblue}{0.137 0.298 0.513}
\newrgbcolor{verde}{0.267 0.637 0.492}  
\newrgbcolor{verde2}{0.168 0.582 0.543} 
\newcommand{\fbk}[1]{{\color{fbkblue} #1}}


\def\bra#1{{\langle#1|}}
\def\ket#1{{|#1\rangle}}
\def\bracket#1#2{{\langle#1|#2\rangle}}
\def\inner#1#2{{\langle#1|#2\rangle}}
\def\expect#1{{\langle#1\rangle}}
\def\e{{\rm e}}
\def\proj{{\hat{\cal P}}}
\def\tr{{\rm Tr}}
\def\H{{\hat H}}
\def\Hdag{{\hat H}^\dagger}
\def\Lop{{\cal L}}
\def\Ehat{{\hat E}}
\def\Edag{{\hat E}^\dagger}
\def\Shat{\hat{S}}
\def\Sdag{{\hat S}^\dagger}
\def\Ahat{{\hat A}}
\def\Adag{{\hat A}^\dagger}
\def\U{{\hat U}}
\def\Udag{{\hat U}^\dagger}
\def\Zhat{{\hat Z}}
\def\Phat{{\hat P}}
\def\Op{{\hat O}}
\def\id{{\hat I}}
\def\x{{\hat x}}
\def\P{{\hat P}}
\def\Px{\proj_x}
\def\Pr{\proj_{R}}
\def\Pl{\proj_{L}}
\def\ODR{O_{_{\rm DR}}(t)}
\def\ODRn{O_{_{\rm DR}}(n)}
\newcommand{\equa}[1]{Eq.~(\ref{#1})}


\title{Perturbations and chaos in quantum maps}
            
\author{Dar\'io E. Bullo}     
\author{Diego A. Wisniacki}
\email{wisniacki@df.uba.ar}
\affiliation{\mbox{Departamento de F\'{\i}sica ``J. J. Giambiagi", 
             FCEN, Universidad de Buenos Aires, 1428 Buenos Aires, Argentina}}
\date{\today}

\begin{abstract}
The local density of states (LDOS) is a distribution that characterizes
the effect of perturbations on quantum systems. 
Recently, it was proposed a semiclassical theory for the LDOS of chaotic billiards and maps. 
This theory predicts that the LDOS is a Breit-Wigner
distribution independent of the perturbation strength and also gives a semiclassical expression
for the LDOS witdth.
Here, we test the validity of such an approximation
in quantum maps varying the degree of chaoticity, the region in phase space where the perturbation 
is applying 
and the intensity of
the perturbation. 
We show that for highly chaotic maps or
strong perturbations  the semiclassical theory of the LDOS  is accurate to describe the  quantum distribution.
Moreover, the width of the LDOS is also well represented for its semiclassical expression in the case of mixed
classical dynamics.   
\end{abstract}

\maketitle

\section{Introduction}

\label{sec1}

The response of quantum systems
to external perturbations is a problem of paramount importance
in many areas of physics. 
Many of the properties of complex quantum system change dramatically
when the system is perturbed,  generating
fundamental phenomena as quantum phase transitions, irreversibility
or dissipation. The present development of experimental technics
in quantum complex systems make the understanding and characterization
of the effect of perturbations highly desirable.

The  most likely suitable magnitude to characterize the
effects of perturbations on quantum systems is the local density 
of states (LDOS).  The LDOS, also  called  strength function, 
was introduced by Wigner  \cite{wigner}  to understand 
the statistical properties of the wave 
functions of complex quantum systems.  
The LDOS is the profile of an eigenstate of an
unperturbed quantum system over the eigenbasis of its perturbed
version.  To be more specific, let us consider 
a system with a one parameter dependent Hamiltonian $H(k)$ with eigenfrecuencies $\omega_j(k)$ and eigenstates $|\psi_{j}(k)\rangle$. 
The LDOS of an eigenstate $|\psi_{i}(k_0)\rangle$ (that we call unperturbed) is given by
\begin{equation}
\rho_{i}( \omega,\delta k)= \sum_{j} \vert\langle\psi_{j}(k)\vert\psi_{i}(k_{0})\rangle\vert^{2}\delta(\omega-\omega_{ij}(k,k_0)) , 
\label{ldos}
\end{equation}
with $\omega_{ij}(k,k_0)=\omega_i(k_0)-\omega_j(k)$ and $\delta k \equiv k-k_0$ the perturbation strength. Eq. \ref{ldos} shows that the LDOS is a density of states 
in which the delta functions are weighed by the overlaps
between perturbed and unperturbed states. In addition,  the LDOS width gives an estimation of how many perturbed states contribute to an unperturbed
one. Besides, it is the Fourier transform  of  the fidelity amplitude (FA) of 
the state $|\psi_{i}(k_0)\rangle$,
\begin{equation}
\rho_{i}( \omega,\delta k)= {\cal{F} } [ \langle\psi_{i}(k_0)\vert e^{i H(k) t/\hbar}e^{ -i H(k_0) t/\hbar} \vert  \psi_{i}(k_{0})\rangle].
\end{equation}
Both the FA and its absolute square value, called the Loschmidt echo, are important measures of sensitivity to 
perturbations and irreversibility of quantum evolutions \cite{jalabert,prosen-rev,jacquod,diego,scholarpedia}.

The LDOS has been considered in many contexts.  In a seminal paper, Wigner studied the LDOS 
in a simple model of banded random matrices \cite{wigner}. 
Subsequently, many authors have used the LDOS to characterize the structure of the eigenstates of 
different random matrix models  \cite{casati1, casati2, jacquod1}. The LDOS has also been studied  in several 
microscopic systems as for example in a Ce atom \cite{Flambaun1}, in chaotic billiards \cite{Doron1} 
or a system of  a particle that evolves in a smooth Hamiltonian  \cite{cohen2}.
In addition, the LDOS has been studied to characterize the effect of perturbations 
in the operation of quantum computers in the presence of static imperfections \cite{georgeot,casati3}. 
It was shown that depending on the characteristics of the system, the LDOS has many regimes as a function 
of the perturbation strength $\delta k$. However, all the mentioned studies have revealed a region of 
perturbation strength in 
which the LDOS has a lorentzian shape, that has been usually called Breit-Wigner distribution.

A step forward has been recently made in the understanding of the LDOS for chaotic systems \cite{Natalia}.  
Its  relation  with the  FA has been exploited to
develop a semiclassical theory of the LDOS for locally perturbed billiards or maps, that is, when 
the perturbation is 
concentrated in a small region of the phase space accessible for the system. It was shown that the LDOS 
has a Lorentzian shape 
under very general perturbations of arbitrarily high intensity an a semiclassical expression 
for its width was derived. 
This expression only depends on the perturbation, while the properties of the system 
are taken into account through a uniform measure in phase space. The same results
were obtained in a subsequent publication for maps that are globally perturbed but the dynamics was assumed to be  
completely random \cite{Nacho1}. 

The aim of our study is to  test  the validity of the semiclassical theory of Ref. \cite{Natalia, Nacho1} in quantum maps when the perturbation is
applied in all the phase space and the dynamics of the classical map is not completely random. We also consider
perturbations that act in an  region of the phase space.
We study  the behavior of the LDOS
for maps with different  degree of chaoticity and intensity of the perturbation. For this purpose we consider two of the most paradigmatic 
systems of quantum chaos studies: the perturbed cat map and the Harper map.    We show that the semiclassical approximation of the
width of the LDOS works very well even for systems with mixed dynamics in which chaos coexist with regular islands. 
The prediction of Lorenzian shape of the LDOS is fulfilled for highly chaotic maps or when the intensity of the perturbation is
big enough. 

The paper is organized as follows. In Sec. \ref{secmaps} we introduce the dynamical systems that we have used for the
numerical study, the 
 cat and the Harper maps.  In this section,we  describe the main characteristics of the classical and quantum dynamics of the maps.
 Sec. \ref{sec2}  is devoted to present the semiclassical theory of
the LDOS  \cite{Natalia, Nacho1}.  
The starting point of this theory is a semiclassical approximation of the fidelity amplitude called  Dephasing Representation \cite{vanicek}. In Sec. \ref{results} we study the behavior of the LDOS for the systems in several situations and test the validity of the semiclassical theory. We consider various degree of chaoticity and intensities of the perturbation. We also compare the cases ol local and global perturbations.
Finally,  we conclude with a summary of our results and some final remarks in Sec. \ref{conclu}.

\section{Systems: maps on a torus}

\label{secmaps}

An usual procedure to understand a complex behavior is to consider very simple systems
in which such a phenomena is observed. The most simple dynamical systems which develops 
all types of complexity  are abstract maps. Due to their simplicity, classical and quantum maps have been 
very important in the development of classical and quantum chaos 
\cite{Hannay, Balazs-Voros, Keating}. 
Furthermore, many quantum maps have been implemented experimentally in previous studies
\cite{Exp 1,Exp 2, Exp 3}.
 
In this paper we have used maps acting on a torus phase space of area ${\cal{A} }= 1$.  In particular 
we have considered the well known cat and Harper maps. These maps possess all the
essential ingredients of chaotic and mixed dynamics and are extremely simple from a numerical point
of view.

The cat maps are linear automorphisms of the torus that exhibit hard chaos. Anosov's theorem \cite{Anosov} establishes that the cat maps are structurally stable, that is, 
the orbits of a slightly perturbed map are conjugated to those of the unperturbed map by a homeomorphism. A perturbation of a cat map
can be represented by matrices acting on the coordinates 

\begin{equation}\left[ 
\begin{array}{c}
q^{'} \\
p^{'}
\end{array}
\right] 
 =
G
\left[ 
\begin{array}{c}
q \\
p 
\end{array}
\right]
+
\left[
\begin{array}{c}
0 \\
1 
\end{array}
\right] \epsilon (q, k) \hspace{0.5cm}
\left( mod1\right),\label{Gatos}\end{equation} 
where $G$ is a $2 \times 2$ matrix with integer elements choosen that $Tr \left(G \right)> 2$ and $det(G)=1$ since the maps are
hyperbolic and conservative. We consider a perturbation 

\begin{equation}
\epsilon (q, k) = ( k/2 \pi)[\cos(2\pi q)-\cos(4\pi q)],
\label{catpert}
\end{equation}
 with the perturbation strength $k<0.11$   to satisfy the Anosov theorem \cite{Matos,Anosov}. To take into account different degrees of chaoticity,
 in this paper we have considered the following matrices $G$,

\begin{displaymath}
G_1=
\left(
\begin{array}{c}
2 \\
1 \\
\end{array} 
\begin{array}{c}
1 \\
1 \\
\end{array}
\right)  ,
\;\;\;\;\;\;\;
G_2=
\left(
\begin{array}{c}
80 \\
6399
\end{array}
\begin{array}{c}
1 \\
80
\end{array}
\right).
\end{displaymath}
The corresponding Lyapunov exponents, which determine the rate of exponential divergence of classical trajectories are $\lambda_1\approx 0.96$ and $\lambda_2\approx 5.07$. We note that $\lambda$  is approximately uniform over the whole phase space and nearly independent of $k$ \cite{Ares2}.

Perturbed cat maps do not capture all the possible motions of Hamiltonian systems. The most common situation is 
a mixture of regular islands interspersed by chaotic regions. To consider this general situation the model that we 
have chosen to study is the Harper map in the unit square \cite{Harper},
%
\begin{eqnarray}
q^{'} &=& q- k\sin{2\pi p} \qquad ({\rm mod}\; 1), \nonumber \\
p^{'} &=& p+ k\sin{2\pi q^{'}}  \qquad ({\rm mod}\; 1),
\label{eq:1}
\end{eqnarray}
where $k$ is a parameter that controls the behavior of the system.
This map can be understood as the stroboscopic version of the flow
corresponding to the (kicked) Hamiltonian
%
\begin{equation}
  H(p,q,t) = -\frac{1}{2\pi} \cos(2\pi p) - \frac{k}{2\pi} \cos(2\pi q)
    \sum_n \delta(t-n k).
    \label{eq:2}
\end{equation}
This is an approximated Hamiltonian  for the motion of an an electron in a crystal under the 
action of an external field. 

\begin{figure}[h] 
  \begin{center}
       \includegraphics[width=8cm, angle=0]{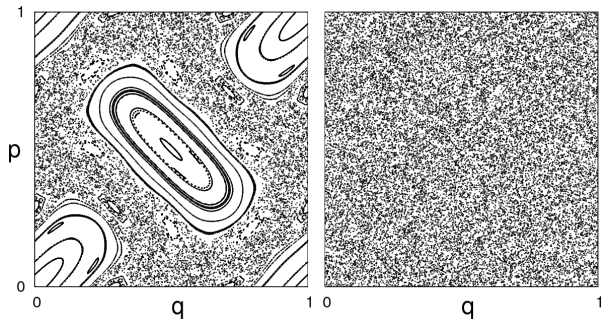} 
       \caption{ \label{fighc} Classical phase space of the Harper map 
   for  $k= 0.3$ (left panel) and $k= 200$ (right panel). See text for details.}
  \end{center}
 
\end{figure}

The Harper map presents a mixed dynamics that depends on the parameter $k$. 
Fig.~\ref{fighc} shows some phase space pictures for this model as an example of the underlying classical dynamics. 
As can be seen in Fig. \ref{fighc} (left panel), 
the system presents a mixed dynamics with regions of regularity
around the origin and the corners coexisting with chaos,
in agreement with the KAM theorem \cite{Anosov}. When the parameter $k=200$ the size of the island are so small
that it is not possible to observe without a finer  resolution [see Fig. \ref{fighc} (right panel)]. 

The quantization on the torus implies  that the wave function should be periodic in both position and momentum 
representation. If in the coordinate and momentum representation the wave function has a period 1 with spacing $1/N$, 
 it follows that $1= 2 \pi \hbar N$. Then, we have a Hilbert space of $N$ dimension for a fixed value of $\hbar$. 
As $N$ takes increasing values, we reach the semiclassical limit. The position basis $\{q_i\}_{i=1}^{N-1}$ (with $q_{i} = i/N$) 
 and momentum basis $\{p_i\}_{i=1}^{N-1}$ (with $p_i = i/N$) are related by the discrete Fourier transform.
 In this setting a quantum map is simply a unitary $U$ 
acting on an $N$ dimensional Hilbert space and evolution after $n$ steps  is given by $U^n$.

There is no general method for map 
quantization. For the perturbed cat map we have considered the quantization based on the  classical 
propagator of Ref.  \cite{Hannay,Matos}. In this case, the matrix elements of the propagator in the position basis are

\begin{eqnarray}
\ U^{C}_{k}(q',q)=\sqrt{\frac{N}{ig_{12}}} \exp \left[ \frac{i\pi N}{g_{12}}(g_{11}q^{2}-2q'q+g_{22}q'^{2}) \right]
\nonumber\\
\exp \left[ \frac{i k N }{2\pi}(\sin(2\pi q)-\frac{1}{2}\sin(4\pi q)) \right],
\label{pertq1}
\end{eqnarray}
where $g_{i,j}$ are the elements of the matrix $G$ and we have used $g_{12}=1$. 

For the Harper map \cite{Harper},  the matrix elements of evolution operator in the mixed basis of position and momenta are

\begin{equation}
\label{eq:harp}
U^{H}_{k}(q,p)=e^{i N k \cos (2 \pi q)} e^{i N k \cos (2 \pi p)}.
\end{equation}

\section{Semiclassical theory of the LDOS of chaotic maps}
\label{sec2}

The LDOS as defined in Eq. \ref{ldos} depends on the characteristics of the state $|\psi_i(k_0)\rangle$. To avoid  
any dependence with some particular  characteristics of this state 
an average over unperturbed states is performed. Due to the finite number of states in quantum maps, we
average over all the Hilbert space.   Thus, the averaged LDOS $\rho (\omega,\delta k)$ is

\begin{equation}
 \rho (\omega,\delta k) =\frac{1}{N}\sum_{i=1}^{N} \rho_{i} (\omega,\delta k) . \label{LDOSA}
\end{equation}

The inverse Fourier transform of Eq. \ref{LDOSA}, the so called average fidelity amplitude (AFA), 
is the starting point of the a semiclassical approximation of the LDOS,

\begin{eqnarray}
 \overline{ O(t,\delta k)  } =\frac{1}{N} \sum_{i} \langle \psi_i(k_0) | e^{iH(k)t/\hbar} e^{-iH_0(k_0)t/\hbar} | \psi_i(k_0) 
\rangle . 
\label{transformada}
\end{eqnarray}

To evaluate   Eq. \ref{transformada} we have used the so called dephasing representation,  a semiclassical 
formulation for fidelity amplitude  which avoids 
the usual trajectory-search problem of the standard semiclassics \cite{vanicek}. One of the forms of the FA obtained 
using the dephasing representation  is 

\begin{eqnarray}
 O_{\phi}(t,\delta k) =  \int W_{\phi}(q,p) e^{-i\Delta S_{t}(q,p,\delta k)/\hbar} dqdp ,
\end{eqnarray}
where $\Delta S_{t}(q,p,\delta k)$ is the action difference evaluated along the umperturbed orbit starting at $(q,p)$ that evolves 
at a time $t$ and $W_{\phi}(q,p)$ is the Wigner function of the initial state $|\phi \rangle$.  Then, 
\begin{eqnarray}
\overline{O(t,\delta k)} =  \int W(q,p) e^{-i\Delta S_{t}(q,p,\delta k)/\hbar} dqdp ,
\end{eqnarray}
where $W(q,p)= (1/N) \sum W_i(q,p) $, with $W_i(q,p)$ being 
the Wigner function of  $\vert \psi_i (k_0)\rangle$.
For chaotic systems, the mean value of the Wigner function for a base of eigenstates is approximately a uniform distribution so $W(q,p)=1/V$
where $V$ is the volume of the phase space.  
Therefore,
\begin{equation}
 \overline{ O(t, \delta k) } = \frac{1}{V} \int e^{-i\Delta S_{t}(q,p,\delta k)/\hbar} dqdp \label{O(t)semi}.
\end{equation} 

Time is discrete in maps, so from now on we use the integer $n$ to count time steps and $V$ is the area of the phase space
that in our case is equal to unity.

In order to solve Eq. \ref{O(t)semi} for maps we need to assume that trajectories become uncorrelated between two successive 
hits in the  perturbed region. This approximation is valid when the perturbation acts on an infinitesimal portion of phase space 
\cite{Natalia,Goussev,Nacho2} or if the unperturbed
dynamics of the system is completely random \cite{Nacho1}. 

Here we have considered the second case,  the $\lambda \rightarrow \infty$ limit, by assuming that the dynamics is purely random. 
This evolution is completely stochastic in the sense that there is no correlation for the different times of the evolution. 
Then, to compute  $\overline{ O(n, \delta k) }$, we have divided  the phase space in $N_c$ cells. The probability to jump 
from cell to any other in phase space is uniform. Therefore it is straightforward to show that the mean FA results

\begin{eqnarray}
 \overline{ O(n,\delta k) } &=& \sum_{j_1} ... \sum_{j_n} e^{[-i(\Delta S_{j_1} + ... + \Delta S_{j_n})/\hbar]}
\nonumber\\
&=& \left( \sum_{j} e^{(-i \Delta S_{j}/\hbar)} \right) ^n  ,
\end{eqnarray}
where $\Delta S_{j p}$ is the action difference evaluated in the cell $j$ at time $p$. The continuous limit is approached when 
$N_c\rightarrow \infty$ resulting in

\begin{equation}
\overline{ O(n,\delta k) } = \left( \int e^{-i \Delta S (q,p,\delta k) /\hbar } dqdp \right)^n , \label{integral}
\end{equation}
where $\Delta S(q,p,\delta k)$ is the action difference after one step of the map.

The exponential decay of Ec. \ref{integral} can be rewritten as 

\begin{equation}
 \overline{ O(n,\delta k) } = e^{- \Gamma n +i \varphi n } \label{n},
 \end{equation}
with
\begin{eqnarray}
  \Gamma  &=& -\ln(|\int e^{-i \Delta S (q,p,\delta k)/\hbar} dq dp| )   \label{1} .
\end{eqnarray}
and  

\begin{eqnarray}
 \varphi  &=& \arg(\int e^{-i \Delta S (q,p,\delta k)/\hbar} dq dp)   \label{2} .
\end{eqnarray}
We note that $\Gamma$ and $\varphi$ depend on the perturbation strength $\delta k$.




Now, we obtain the semiclassical expression for the average LDOS by the inverse Fourier transform of Eq. \ref{n},

\begin{equation}
 \rho_{sc}(\omega,\delta k) = {\cal{F}}^{-1}_{[\bar{O}]}(\omega,\Gamma,\varphi) = \frac{\Gamma }{\pi[(\omega-\varphi)^2 +
\Gamma^2]}  .
\end{equation}
The phase $\varphi$ determines the location of the center of the Lorenzian function and $\Gamma$ its width.

Finally, we have to take into account the fact that the spectrum of a map is periodic because of a compact phase space. This periodicity changes the 
form of the LDOS into a periodized Lorentzian function

\begin{eqnarray}
 {\rho}_{sc}(\omega,\delta k) &=&  L^{(p)}(\omega,\Gamma,\varphi)  
\nonumber\\
&=& \sum _{j=-\infty} ^{\infty} \frac{\Gamma}
{\pi[(\omega -\varphi- 2\pi j)^2+\Gamma^2]} .  \label{LorPeriod}
\end{eqnarray}

The same  semiclassical expressions for the  LDOS were obtained in Ref. \cite{Natalia,Nacho2} when the 
perturbation acts in a region of the phase space of area $\alpha \rightarrow 0$.

\begin{figure} 
  \begin{center}
       \includegraphics[width=9.0cm, angle=0]{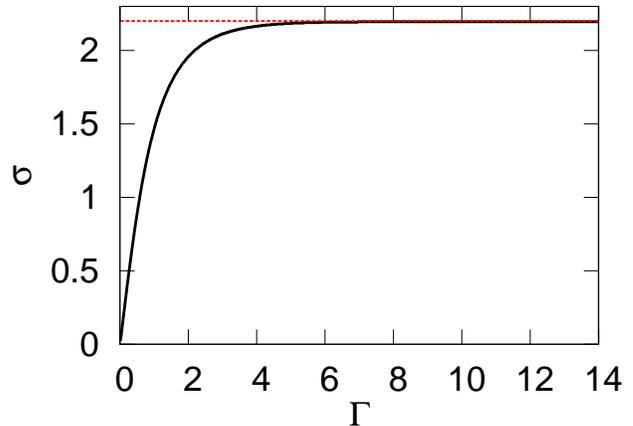} 
        \caption{\label{FASE1}\footnotesize (Color online) .  Witdh $\sigma$ vs. $\Gamma$ for a periodized Lorentzian function 
        [Eq. \ref{LorPeriod}]. The limit of $\sigma$ for 
    $\Gamma \rightarrow \infty$ which corresponds for a constant  LDOS   is also plotted  with red dotted line. }
  \end{center}
\end{figure}

A magnitude that has physical interest is the width $\sigma$ of the LDOS which  is a measure of the number of perturbed states that are 
needed to describe a unperturbed one. Therefore, this quantity offers clear information  about the effect of perturbations on 
on a quantum system. Moreover,  the width of the LDOS determines for some regime of the perturbation,  the rate of fidelity decay under
imperfect motion reversal (the Loschmidt echo). There are different ways of determining this width of a distribution. 
In our case we are going to take the distance about the  average value of the LDOS that contains 70 \% of the probability. That is,
\begin{equation}
  \int^{\langle \omega \rangle + \sigma}_{\langle \omega \rangle - \sigma} \rho(\omega, \delta k) d \omega = 0.7 \hspace{0.4cm} . \label{ANCH}
\end{equation}
where 

\begin{equation}
 \langle \omega \rangle= \int^{\pi}_{-\pi} w \rho(\omega, \delta k) d \omega 
\end{equation}

We show in Fig. \ref{FASE1} the relation between its width $\sigma$ 
and $\Gamma$ for the periodized Lorentzian function of Eq.  \ref{LorPeriod} .


\section{Results}
\label{results}

\begin{figure} 
  \begin{center}
       \includegraphics[width=9.0cm, angle=0]{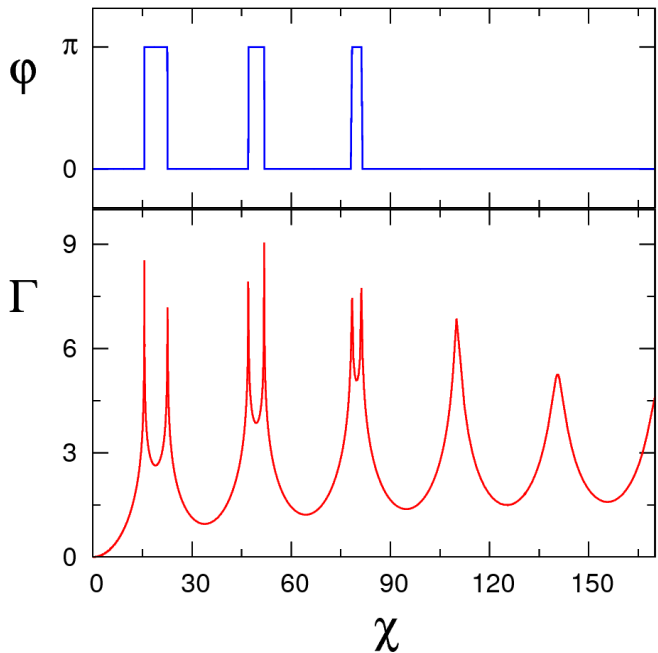} 
        \caption{\label{FASEGATO}\footnotesize (Color online) $\Gamma$
and the phase $\varphi$ as a function of the scaled pertubation $\chi$ for the perturbed cat map. }
  \end{center}
\end{figure}
The main interest of a semiclassical theory is to describe quantum mechanical quantities using classical information. 
In this section we show the behavior of the LDOS for the quantum maps presented before and  test the validity of the 
semiclassical approximation of the LDOS described in the previous section.
The aim this section is to compare the approximated $\rho_{sc}$ and $\sigma_{sc}$ with the   
corresponding exact quantum values. The latter are numerically computed by diagonalization of the
evolution operators of Eq. \ref{pertq1} and \ref{eq:harp}.
 
The semiclassical approximation of the LDOS is completely determined by $\Gamma$ and $\varphi$ that are obtained
with the calculation of the integral of Eq. \ref{integral}.  To avoid the dependence of the results with the dimension of 
the Hilbert space $N$ we have considered all the studied quantities as a function of 
the scaled strength of the perturbation  
\begin{equation}
\chi \equiv (k-k_{0})/(2 \pi \hbar) =\delta k N.
\end{equation} 
In all the calculations included in this section the number of states of the Hilbert space set as $N=2000$.

\subsection{Peturbed cat map}
The action difference for one iteration of the perturbed cat map described in Sec. \ref{secmaps} is given by

\begin{equation}
 \Delta S (q,p,\delta k)= \left( \frac{\delta k}{4 \pi^2} \right) \left[  \sin(2\pi q)-\frac{1}{2} \sin(4 \pi q)  \right]  . \label{ACCIONGATO}
\end{equation}
Using  Eq. \ref{ACCIONGATO} , \ref{1} and \ref{2} we compute
$\Gamma$ and $\varphi$. In Fig. \ref{FASEGATO} we plot $\Gamma$ and $\varphi$ 
for the perturbed cat map as a  function of the scaled perturbation strength $\chi$. We can see that for perturbation of Eq. \ref{catpert}, $\varphi$  
has only two possible values either $0$ or $\pi$.

\begin{figure} 
  \begin{center}
       \includegraphics[width=9cm, angle=0]{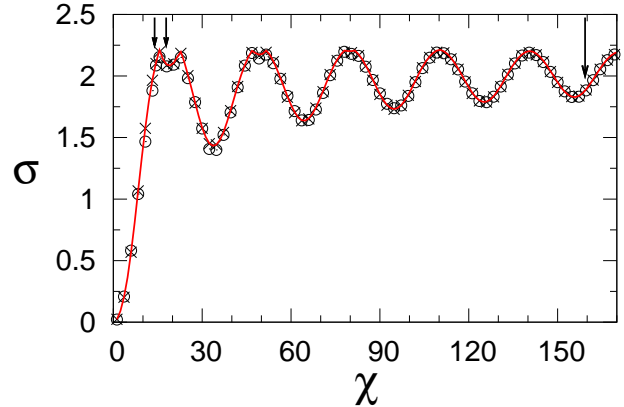} 
 \caption{\label{ANCHO}\footnotesize (Color online) Width $\sigma$ of the LDOS as a function of the scaled 
perturbation strength $\chi=(k-k_0) N$ for the cat map with $G_1$ ({\large $\circ$}) and $G_2$  ({\large $\times$}).   
The red solid line is the semiclassical approximation of $\sigma$. The number of states of the Hilbert space   $N=2000$ and $k_0=0.01$.   We indicate with arrows
the perturbations strength of the LDOS displayed in Fig. \ref{LDOS}. 
   }
  \end{center}
\end{figure}

We firstly compare the semiclassical approximation of the width of the LDOS with the corresponding quantum value. 
For this reason  the  width of the LDOS has been computed for the cat map using $k_0=0.01$ to avoid all the arithmetic peculiarities 
of the cat map ($k=0$), which account for the non-generic spectral statistics \cite{Keating2}.  
In Fig. \ref{ANCHO}  the width of the LDOS is shown for the cat maps with  $G_1$ and $G_2$. The semiclassical approximation  $\sigma_{sc}$, plotted
in solid line,  works extremely well for both 
cat maps in the whole  range of considered perturbations.  

The width of the LDOS  $\sigma$ for the cat maps has two clearly different regimes [ Fig. \ref{ANCHO}] . 
For small perturbation strength  when $\chi \lesssim 10$ it presents a quadratic behavior that is usually  
called Fermi Golden Rule regime. Conversely for greater strength the width is an oscillating function. In order to understand the behavior of
$\sigma_{sc}$ when $\chi \rightarrow \infty$  we have used the stationary phase approximation 
method to solve the integral Eq. \ref{O(t)semi} obtaining

\begin{displaymath}
 \Gamma \rightarrow -\log[1/\sqrt{\chi}]  \hspace{0.5cm} for \hspace{0.3cm} 
 \chi  \rightarrow \infty	,
\end{displaymath}
therefore the width $\sigma_{sc} \rightarrow 0.7 \pi$ that corresponds to a uniform distribution. 

\begin{figure} 
  \begin{center}
       \includegraphics[width=8.5cm, angle=0]{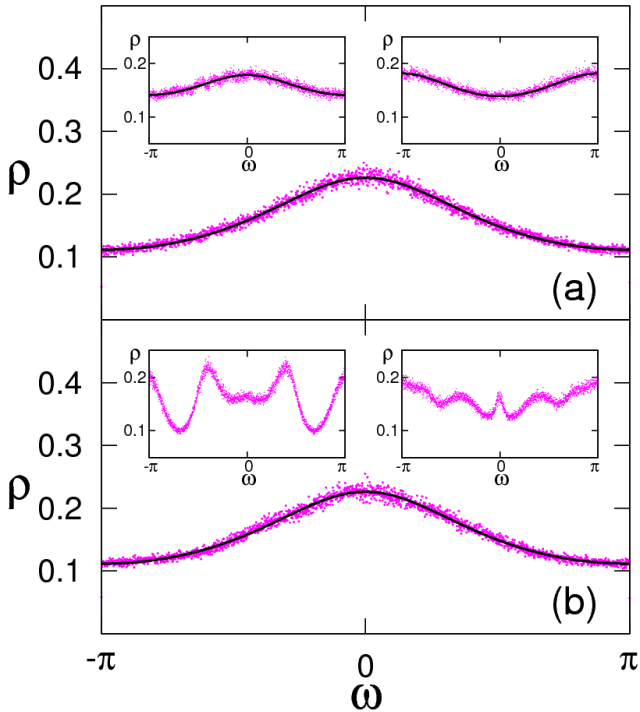} 
        \caption{\label{LDOS}\footnotesize (Color online) LDOS $\rho$ (points) and its semiclassical approximation (solid 
line). (a) cat map of Eq. \ref{Gatos} with $G=G_2$ for a scaled perturbation strengh $\chi=159.6$ (main plot), $\chi=14.4$ (left inset) and  
$\chi=18.0$ (right inset).
(b)  cat map of Eq. \ref{Gatos} with $G=G_1$  for  $\chi=159.6$,  $\chi=14.4$ (left inset) and  
$\chi=18.0$ (right inset).}
  \end{center}
\end{figure}

At this point we would like to see how good the semiclassical approximation of the LDOS can describe the complete distribution. 
We have therefore computde the quantum LDOS for several values of perturbation strength for the cat maps $G_1$ and $G_2.$
In Fig. \ref{LDOS} we compare the LDOS with its semiclassical approximation for perturbations indicated in Fig. \ref{ANCHO}
with arrows. 
Fig.  \ref{LDOS}(a)  corresponds to the most chaotic case $G_2$. 
In the main plot $\chi=159.6$, in the left inset $\chi=14.4$ and in the right inset $\chi=18$. 
We can see that the semiclassical approximation works very well
for all the perturbations, that is, the LDOS is a periodized Lorentzian function indistinctly of the  perturbation strength.
Left and right inset corresponds to approximately 
the same width of the distribution but in the left figure $\varphi=0$ and $\varphi=\pi$ for the right so in this case the periodized Lorentzian is
centered in $ \omega=\pi$. 
As can be seen in Fig. \ref{FASEGATO}, near $\chi \approx 15$, the phase  $\varphi$ has a discontinuity and jumps
from $0$ to $\pi$, for this reason the center of the LDOS changes from $\omega=0$ to $\omega=\pi$. Similar behavior occurs
in the other discontinuities of $\varphi$ near  $\chi \approx 50$ and $70$.

\begin{figure} 
  \begin{center}
       \includegraphics[width=8.0cm, angle=0]{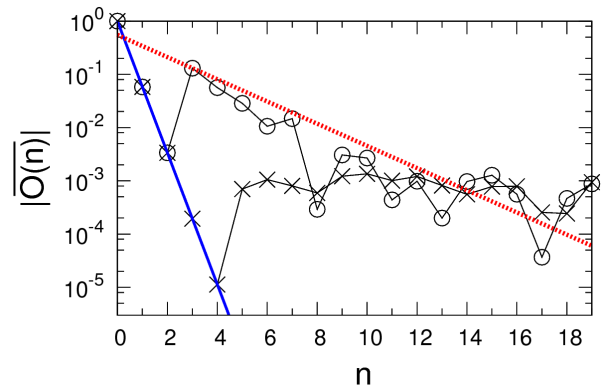} 
       \caption{ \label{odet}(Color online)  Mean value of the amplitude fidelity $\overline{O(n)}$ for the cat map with $G_1$ ({\large $\circ$}) and 
$G_2$ ({\large $\times$}) with perturbation strength $\chi=14.4$. The exponential decay given by $\exp(-\Gamma n)$ is plotted  with solid blue line.
We also plot  with dotted red line  the exponential decay given by $\exp(-\lambda n/2)$ with $\lambda$ the Lyapunov exponent for the cat map with $G_1$.  }        
  \end{center}
\end{figure}

In  Fig. \ref{LDOS}(b)  the results for the cat map with the matrix $G_1$ are shown. In the figure main panel, we show that for big
 perturbation strength after the quadratic regime ($\chi=159.6$) the LDOS is well described by the semiclassical Lorentzian distribution. 
Conversely for smaller perturbations strength  the LDOS does not show a Lorentzian behavior [see inset of  Fig.  \ref{LDOS}(b)].
 To understand this behavior we show in Fig. \ref{odet} 
the mean value of the fidelity amplitude $\overline{O(n)}$ for $\chi=14.4$ of both cat maps with $G_1$ ({\large $\circ$}) and 
$G_2$ ({\large $\times$}) which corresponds to the inverse Fourier transform of the LDOS plotted in the left inset of Fig. \ref{LDOS}(a)
and (b). We see that in the case in which the LDOS is not a periodized Lorentzian function the corresponding $\overline{O(n)}$ has a big revival
(at $n=4$). This kind of behavior, known as survival collapse after which the largest revivals appear,  was observed in a spin chain \cite{revivals} 
and can be the cause  for non-Markovian quantum evolutions \cite{non-markovian}.

\begin{figure} 
  \begin{center}
       \includegraphics[width=8.0cm, angle=0]{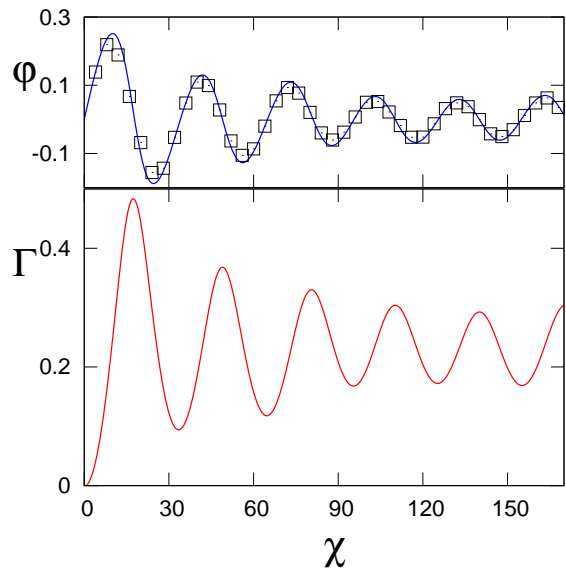} 
       \caption{\label{FASEGATOCORTADO}\footnotesize (Color online) $\Gamma$  
and the phase $\varphi$  as a function of the scaled pertubation $\chi$ for the cat map when the perturbation is
applied  in a $q$ strip from $q_0=0.25$ to $q_1=0.46$. The mean value of the quantum LDOS is also plotted with ($\Box$).   }
  \end{center}
\end{figure}

We test now the validity of the semiclassical approximation of the LDOS for local perturbation. For this reason the perturbation is  applied 
in a $q$ strip from $q_0=0.25$ to $q_1=0.46$   so the area of the perturbed region 
is $\alpha=\Delta q \Delta p=q_1-q_0=0.21$. In Fig. \ref{FASEGATOCORTADO} we show $\Gamma$ and $\varphi$ as a function of the 
scaled perturbation strength $\xi$ computed using  Eq. \ref{1} and \ref{2}.
We can see that for this local perturbation  $\varphi$ is an oscillating function so the semiclassical approximation of the LDOS 
is periodized Lorentzian function with an oscillating mean value. In Fig. \ref{FASEGATOCORTADO}(top panel) the mean value of the exact LDOS   it is also plotted with 
({\large $\Box$}) showing that the semiclassical $\varphi$ describe very well this quantity.

\begin{figure} 
  \begin{center}
       \includegraphics[width=9.5cm, angle=0]{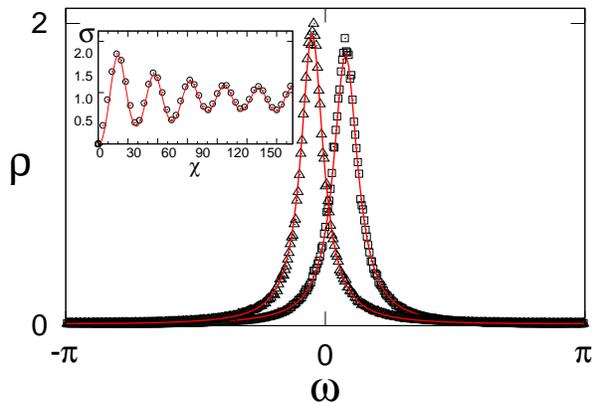} 
        \caption{\label{fig-ldos-cut}\footnotesize (Color online) LDOS $\rho$ for a local perturbations. The cat map with $G_1$ is perturbed 
        from $q_0=0.25$ to $q_1=0.46$. The scaled perturbation strength $\chi=8$ ($\bigtriangleup$) and $\chi=28$ ($\Box$). The semiclassical
        approximation of the LDOS is plotted with red solid line. Inset: Width $\sigma$ of the LDOS as a function of the scaled 
perturbation strength $\chi$ ($\circ$) and with red solid line the semiclassical approximation.}
  \end{center}
\end{figure}

The LDOS is also very well approximated by the semiclassical LDOS for all the perturbations strength that we have studied.
In Fig. \ref{fig-ldos-cut} we show the LDOS for in $\chi=8$ when the width grows cuadratically (FGR regime) and for $\chi=28$ when   
the width shows an oscillating behavior. The semiclassical approximation is plotted with solid line.  
In the inset of Fig  \ref{fig-ldos-cut}  we show the width of the LDOS for this local perturbation
and its semiclassical approximation. We can clearly see that the $\sigma_{sc}$ works very well for local perturbation. It is noteworthy
that all the calculations for local perturbations were done using the map $G_1$ showing that when the perturbation is applying is a
small region of the phase space less degree of chaoticity is needed for the semiclassical LDOS to be accurate.

\subsection{Harper map}

We have studied  the LDOS of the Harper map using the evolution operator of Eq. \ref{eq:harp} with $k=k_0+ \delta k$. 
The parameter $\delta k$ is the perturbation strength and  as we have used for the cat map,  the scaled perturbation straight
$\chi=\delta k N$. In this case the action difference for one iteration of the 
Harper map is given by
\begin{equation}
 \Delta S (q,p,\delta k)= \left( \frac{\delta k}{2 \pi} \right) \left[  \cos(2\pi p)+ \cos(2 \pi q')  \right]  .  \label{ACCIONHARPER}
\end{equation}
where $q'$ is given by Eq. \ref{eq:1}. 

\begin{figure} 
  \begin{center}
       \includegraphics[width=9.0cm, angle=0]{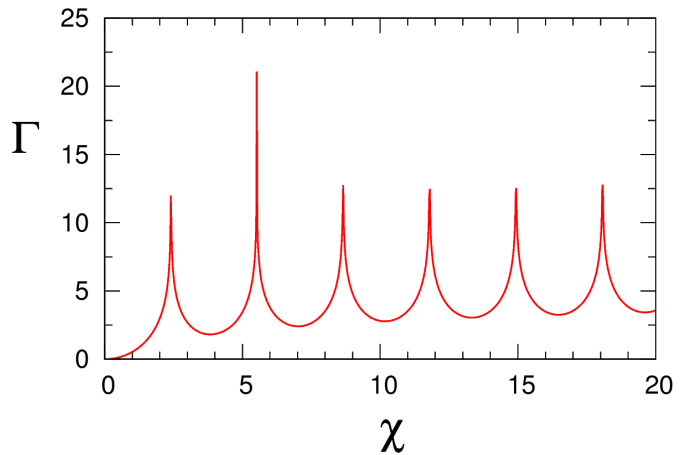} 
        \caption{\label{GAMAHARPER}\footnotesize (Color online) $\Gamma$ as a function of the scaled pertubation strength $\chi$ for the Harper map.  }
  \end{center}
\end{figure}

We have  considered as unperturbed system the cases with  $k_0=0.30$ [mixed dynamics, Fig. \ref{fighc}(left panel)] and 
$k_0=200$ [chaotic dynamics, Fig. \ref{fighc}(right panel)].  
Using Eq. \ref{1}, \ref{2} and \ref{LorPeriod}  we compute $\Gamma$, $\varphi$  
and the corresponding semiclassical approximation of the LDOS.  In Fig. \ref{GAMAHARPER} we show $\Gamma$
as a function of the scaled perturbation strength $\chi$. For the action difference of the Harper map [Eq. \ref{ACCIONHARPER}] we have
obtained  that $\varphi=0$.

\begin{figure} 
  \begin{center}
       \includegraphics[width=9cm, angle=0]{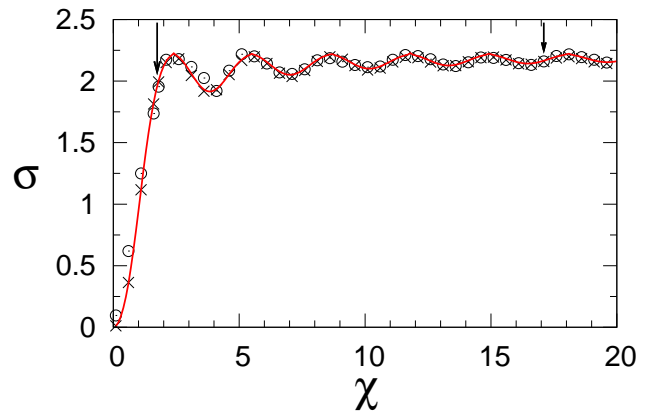} 
     \end{center}
     \caption{(Color online) Width $\sigma$ of the LDOS as a function of the scaled 
perturbation strength $\chi$ for the Harper map with $k_0=0.3$ ({\large $\circ$}) and $k_0=200$ ({\large $\times$}).
In red line it is plotted the semiclassical approximation $\sigma_{sc}$.  We indicate with arrows
the perturbations strength of the LDOS displayed in Fig. \ref{LDOSHARPER}
}
  \label{HARPERWIDTH}
\end{figure}

In Fig. \ref{HARPERWIDTH} we show the width of the LDOS for the Harper map and the corresponding semiclassical
approximation.  When the dynamics of the Harper map is completely chaotic, the semiclassical $\sigma_{sc}$ works well as
expected. Surprisingly,  the semiclassical approximation works reasonably well even for mixed dynamics. 
This agreement is more noticeable for bigger $\chi$. 
The explanation of this unexpected behavior is as follows.  Eq. \ref{integral} is exact
for one time step ($n=1$)  and if the perturbation strength is big enough the fidelity amplitude decays  in this short time. Therefore, this
short time decay gives the width of the Fourier transform which is the LDOS. 

\begin{figure} 
  \begin{center}
       \includegraphics[width=8.5cm, angle=0]{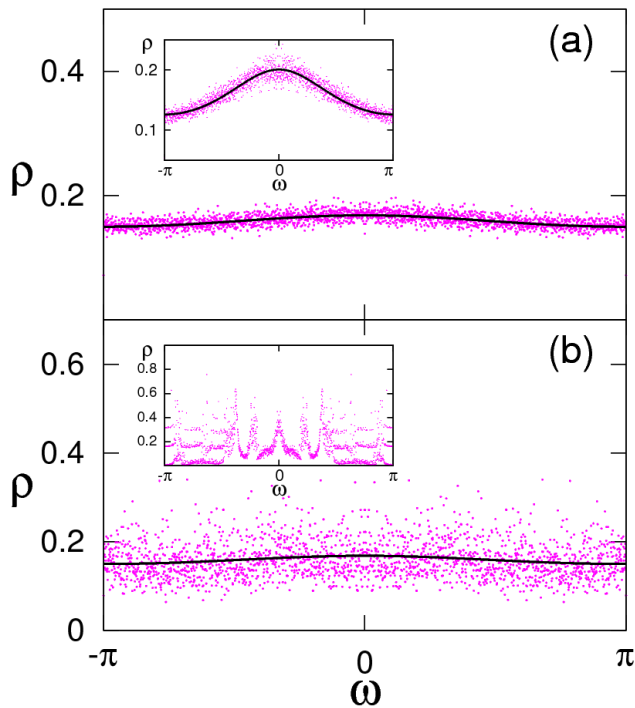} 
      \caption{\label{LDOSHARPER}(Color online) 
 $\rho$ (points) and its semiclassical approximation $\rho_{sc}$ (solid line) for the Harper map.  
(a)  $k_0=200$. In the main plot  $\chi=17$ and
in the inset $\chi=1.8$. In both plot it is  seen that the semiclassical LDOS describes 
the full quantum result.
(b) $k_0=0.3$ (mixed dynamics). In the main plot  $\chi=17$ and
in the inset $\chi=1.8$.}
  \end{center}
\end{figure}

In Fig. \ref{LDOSHARPER} we show the LDOS for the Harper map. 
Although the semiclassical width of the LDOS $\sigma_{sc}$ works well for mixed dynamics, the complete distribution is not  well 
reproduced by a periodized Lorentzian distribution. This is shown in the inset of Fig. \ref{LDOSHARPER} (b) for the Harper map with $k_0=0.3$ 
and $\chi=1.7$. If the perturbation strength is bigger [Fig. \ref{LDOSHARPER} (a) (main plot)] the  semiclassical theory works reasonably well
but the quantum LDOS is a more fluctuating function than the chaotic case [see Fig. \ref{LDOSHARPER} (a)] .
As expected,   the semiclassical theory works well for the case of $k_0=200$ in which the Harper map is fully chaotic (Fig. \ref{LDOSHARPER} (a)).

\section{Conclusions}
\label{conclu}

The reaction of a system to perturbations is a fundamental problem in quantum mechanics. In this paper  we have made a detailed analysis of the 
response to perturbations of the simplest  quantum systems, which can have complex classical dynamics. 
For this reason, we have studied the LDOS in  the perturbed cat map, a completely chaotic system
and the Harper map which has mixed dynamics. Our fundamental goal was to discuss the validity of a semiclassical theory of LDOS that
has been recently developed \cite{Natalia,Nacho1}. This theory is based on the relation of the LDOS with the fidelity amplitude, a measure
of irreversibility and sensitivity to perturbations of quantum systems.  Furthermore,  it uses the dephasing representation of the fidelity amplitude, a semiclassical formulation 
that avoids the usual problems of semiclassical theories. The main assumption of the semiclassical 
theory of LDOS is that the trajectories get uncorrelated after one step of the map. This condition is fullfiled if the dynamics is completely 
random or when the perturbation is applied in a infinitesimal region of the phase space. Due to the fact that these conditions are not achieved in 
dynamical system, we tested the validity of such a semiclassical theory of the LDOS.  

We have analyzed various situations: local and global perturbations and also we have varied the
degree of chaoticity. We show that the LDOS is very well described by its semiclassical expression when the map is highly chaotic, either if 
the perturbation is localized in phase space, or when the perturbation strength is big enough . We remark that in these cases the 
semiclassical LDOS completely reproduces the quantum version
without any fiting parameters.
We have studied the case of mixed dynamics and surprisingly enough our results show that the semiclassical width of the LDOS describes the 
full quantum version even in this case. 
 
We would like to highlite that our results could be of importance in the study of the LDOS of billiards. Indeed,  the behavior of a billiard system has 
many resemblances  with maps. For example, the classical dynamics of a billiard can be described by a map on the boundary.
Quantum billiards are realistic systems that can be constructed in experimental setups of several nature. In fact, there are 
cavities of microwave, acoustic or optical wave.  The semiclassical approximation of
the width of the LDOS has been successfully applied in billiard that has been perturbed both locally \cite{Goussev, Goussev2} and 
globally \cite{Natalia}. However, in these works the behavior of the whole distribution was not properly discussed. 
Further insight on the LDOS of this systems will be part of future studies. 

\section{Acknowledgements}

The authors acknowledge the support from CONICET (PIP-6137) , UBACyT (X237, 20020100100741, 20020100100483) 
and ANPCyT (1556).  We would like to thank Ignacio Garc\'{\i}a Mata for useful discussions.

\end{document}